%% file: mgw_j_stat-mech24.tex
\documentclass[12pt]{iopart}

\usepackage{graphics}

\usepackage[utf8]{inputenc}
\usepackage[T1]{fontenc}

\usepackage{mathtools}
\usepackage{bm}
\usepackage{siunitx}
\usepackage{booktabs}
\usepackage{csquotes}

\input{macros.tex}

\usepackage{hyperref}
\newcommand{\creflastconjunction}{, and\nobreakspace}

\expandafter\let\csname equation*\endcsname\relax

\expandafter\let\csname endequation*\endcsname\relax

\begin{document}

\title{Exact solution of the nonlinear boson diffusion equation for gluon scattering}

\author{L.\,M\"ohringer and G.\,Wolschin}

\address{Institut f\"ur Theoretische Physik der Universit\"at Heidelberg, Philosophenweg 16, D-69120 Heidelberg, Germany, EU}
\vspace{10pt}

\begin{abstract}
An exact analytical solution of the nonlinear boson diffusion equation (NBDE) is presented. It accounts for the time evolution towards the Bose--Einstein equilibrium distribution through inelastic and elastic collisions in case of constant transport coefficients. As a currently interesting application, gluon scattering in relativistic heavy-ion collisions is investigated. An estimate of time-dependent gluon-condensate formation in overoccupied systems through number-conserving elastic scatterings in Pb-Pb collisions at relativistic energies is given.
\end{abstract}

\section{Introduction}
\label{intro}

To account for the time evolution of the fast thermalization of partons in the initial stages of relativistic heavy-ion collisions \cite{hot23} towards local statistical equilibrium, a nonlinear diffusion equation for the occupation-number distributions in the full momentum range has been proposed \cite{gw18,gw22} as an approximation to the quantum Boltzmann equation.
Whereas the latter and other approximate kinetic equations \cite{jpb12,blmt17} that are using the small-angle approximation for gluon scattering \cite{jpb13} must be solved numerically, the nonlinear diffusion equation can be solved analytically in the limit of constant transport coefficients, and thus provides a transparent and analytically tractable model for aspects of the kinetic evolution of a dense parton system during the early moments of a relativistic heavy-ion collision.
 
 Closed-form solutions of the nonlinear equation have been derived for both, fermions (quarks) and bosons (gluons). Although these solutions still depend on the specific form of the initial conditions, it has been possible to derive explicit analytical results for quarks and theta-function initial conditions \cite{bgw19}. For gluons, these are more difficult to obtain, because the boundary condition at the singularity that occurs at the chemical potential $\mu$ must be considered in a combined initial- and boundary-value problem. It has been partially solved for inelastic gluon scattering with $\mu=0$ \cite{gw22}, but the generalized (time-dependent) partition function and its derivative with respect to energy have been computed numerically, although the analytical result for the free case without boundary condition was given earlier \cite{gw18}. We focus on the thermalization of massless gluons with single-particle energy $\epsilon=|\textbf{p}|=p$.

It is the first aim of the present work to analytically derive the exact solution of the nonlinear boson diffusion equation (NBDE) for inelastic gluon scattering including boundary conditions using $\theta$-function initial conditions. The second aim is to extend the model to elastic gluon scattering, and compute schematically time-dependent gluon-condensate formation through number-conserving elastic collisions in overoccupied systems. It is noted, however, that due to the faster timescale of inelastic collisions, transient gluon-condensate formation in relativistic heavy-ion collisions is actually unlikely to occur \cite{blmt17}, although it initially \cite{jpb12} appeared as an intriguing possibility. 

The nonlinear diffusion model is briefly reviewed in the next section. In section\,3, the time-dependent solutions of the NBDE are calculated for $\theta$-function initial conditions. We obtain exact analytical results for the generalized (time-dependent) partition function and its derivative with respect to energy. This allows us to calculate the exact solution of the NBDE through the nonlinear transformation that was found in  Refs. \cite{gw18,gw22}. We investigate the limiting behaviour of the exact solutions for $t\rightarrow 0$ and $t\rightarrow\infty$ in section\,4. The results for the thermalization of the occupation-number distribution function through inelastic collisions are considered in section\,5. Elastic collisions and time-dependent transient gluon-condensate formation in overoccupied systems is discussed in section\,6. The conclusions are drawn in section\,7.

\section{Nonlinear diffusion equation}
\label{model}
The nonlinear partial differential equation
 for the single-particle occupation probability distributions $n\equiv n\,(\epsilon,t)$ had been derived
 from the quantum Boltzmann collision term in \cite{gw18} as
 \begin{equation}
	{\partial_t n}=-{\partial_{\epsilon}}\left[v\, n\,(1\pm n)+n\,{\partial_\epsilon D}\right]+{\partial_{\epsilon\epsilon}}\bigl [D\,n\,\bigr]
\label{nbde}
\end{equation}
where the $+$ sign represents bosons, and the $-$ sign fermions. The drift term $v\,(\epsilon,t)<0$ accounts for the shift of the distribution towards the infrared, the diffusion function $D\,(\epsilon,t)$ for the broadening with increasing time. The derivative-term of the diffusion coefficient is required \cite{gw22}
such that the stationary solutions $n^\pm_\infty(\epsilon)$ become Bose--Einstein or Fermi--Dirac equilibrium distributions, respectively, which are attained for $t\rightarrow\infty$ as
\begin{equation}
n_\infty^\pm(\epsilon)=n_\mathrm{eq}^\pm(\epsilon)=\frac{1}{e^{(\epsilon-\mu)/T}\mp 1}\,,
 \label{Bose--Einstein}
\end{equation}
provided the ratio of drift to diffusion $v/D$ has no energy dependence, and requiring that
$\lim_{t\rightarrow \infty}[-v\,(\epsilon,t)/D\,(\epsilon,t)] \equiv 1/T$. (We use units $k_\text{B}=\hbar=c=1$ throughout this manuscript).
The chemical potential is $\mu\leq0$ in a finite Bose system, and $\mu=\epsilon_\mathrm{f}>0$ for a Fermi system. It appears as an integration constant in the stationary solution of 
Eq.\,(\ref{nbde}) for $t\rightarrow\infty$. It vanishes in case of inelastic collisions that do not conserve particle number such that the temperature alone determines the equilibrium distribution. 

In the present work, we concentrate on the nonlinear diffusion equation for bosons to account for gluon thermalization, and consider both, particle-number conserving elastic collisions, and inelastic collisions that violate particle-number conservation.
We use the ergodic approximation, where the occupation-number distributions 
$n\,(\epsilon,t)$ depend on energy $\epsilon$ and time. 
In relativistic systems, the energy dispersion relation is
$\epsilon=\sqrt{|\textbf{p}|^2+m^2}$ for massive particles, and
$\epsilon=|\textbf{p}|=p$ for massless particles such as gluons. 

The NBDE Eq.\,(\ref{nbde}) can thus be applied to elastic as well as inelastic gluon scatterings,
albeit with differing transport coefficients. In both cases, the detailed properties of the many-body system are hidden in the drift and diffusion functions \cite{gw18,gw22}. In the simplified case of constant transport coefficients, the structure of  the nonlinear diffusion equation \begin{equation}
	{\partial_t n}\,(\epsilon,t)=-v\,{\partial_{\epsilon}}\left[\, n\,(\epsilon,t)(1+n\,(\epsilon,t))\right]+D\,{\partial_{\epsilon\epsilon}}\bigl [n\,(\epsilon,t)\bigr]	
	\label{bose}
\end{equation}
opens up the rare possibility for exact analytical solutions of the nonlinear diffusion problem as considered in the present work,
with the equilibrium temperature $T=-D/v$, and the chemical potential $\mu\le0$ appearing as an integration constant.

The nonlinear diffusion equation not only has the correct equilibrium limit, but also preserves the essential features of quantum statistics that are contained in the quantum Boltzmann equation, such as the Bose enhancement in the low-momentum region that increases rapidly with time. When applied at very low energies to elastic scattering of ultracold bosonic atoms, it properly accounts for thermalization and time-dependent Bose--Einstein condensate formation \cite{gw22a,kgw22} in agreement with deep-quench data for bosonic atoms. For gluons in relativistic collisions, however, this has to be supplemented by inelastic processes that violate particle-number conservation.

The diffusion equation with constant coefficients can be solved in closed form using the nonlinear transformation outlined in \cite{gw18} for any given initial condition $n_\mathrm{i}\,(\epsilon)$. For bosons, one must also consider the boundary conditions occuring at the singularity $\epsilon=\mu\le 0$ \cite{gw22}. No corresponding singularity exists for fermions, such that the fermionic exact solution of the nonlinear problem can be obtained with the free Green's function \cite{gw18,bgw19}. 

The solution method of the nonlinear boson diffusion equation with boundary conditions \cite{gw22} through a nonlinear transformation is first briefly reviewed.
The solution $n\,(\epsilon,t)$ can be written as
\begin{equation}
	n\,(\epsilon,t) =T {\partial_{\epsilon}}\ln{{Z}(\epsilon,t)} -\frac{1}{2}= \frac{T}{{Z}} {\partial_\epsilon{Z}} -\frac{1}{2}
	\label{eq:Nformula}	
\end{equation}
with the time-dependent partition function ${{Z}(\epsilon,t)}$ that fulfills the linear diffusion (heat) equation 
  \begin{equation}
\partial_t\,{Z}(\epsilon,t)=D\, {\partial_{\epsilon\epsilon}}{\,{Z}(\epsilon,t)}\,.
    \label{eq:diffusionequation}
\end{equation}
Without boundary conditions, Green's function \(G_\mathrm{free}(\epsilon , x , t)\) of Eq.\,(\ref{eq:diffusionequation}) is a single Gaussian
\begin{equation}
	G_\mathrm{free}(\epsilon,x,t)=\frac{1}{\sqrt{4\pi Dt}}\,\exp\left[- \frac{(\epsilon-x)^2}{4Dt}\right]\,.
	\label{eq:Greensnonfixed}
\end{equation}
The generalized partition function can then be written as an integral over Green's function $G\,(\epsilon,x,t)$ of the linear diffusion equation
times an exponential function $F(x)=Z(\epsilon,t=0)$
     \begin{equation}
  Z(\epsilon,t)= \int_{-\infty}^{+\infty}{G}\,(\epsilon,x,t)\,F\,(x)\,\text{d}x\,,
    \label{eq:partitionfunctionZ}
    \end{equation} 
where $F$ contains an integral over the initial nonequilibrium gluon distribution $n_\text{i}$ \cite{gw18}
\begin{equation}
	F\,(x) = \exp\,\left[-\frac{1}{2D}\left( v x+2v\int^xn_\text{i}(y)\,\text{d}y \right)\right]\,.
	\label{fini}
\end{equation}
Here, the integration constant in the indefinite integral can be set to zero since it cancels out when taking the logarithmic derivative.\\

 When considering boundary conditions at the singularity $\epsilon=\mu\le 0$ as required for bosons, a
bounded Green's function that equals zero at $\epsilon=\mu~ \forall t$ can be expressed as
\begin{equation}
	G_\text{b}\,(\epsilon,x,t) = G_\mathrm{free}(\epsilon,x+\mu,t) - G_\mathrm{free}(\epsilon,-x+\mu,t)\,.
	\label{Greens}
\end{equation}
 The time-dependent partition function then becomes zero at the singularity for all times,
$Z_\text{b}(\epsilon=\mu,t)=0$, and the occupation-number distribution that is calculated from the nonlinear transformation Eq.\,(\ref{eq:Nformula}) becomes infinity as the energy approaches $\mu$,  \(\lim_{\epsilon \downarrow \mu} n\,(\epsilon,t) = \infty\) \,$\forall$ \(t\). It attains the Bose--Einstein limit over the full energy range as $t\rightarrow \infty$ \cite{gw22}, not only in the thermal tail as in case of the free solutions \cite{gw18}.
The energy range is now restricted to $\epsilon \ge \mu$, and the time-dependent partition function that includes the boundary conditions becomes 
      \begin{equation}
Z_\text{b}(\epsilon,t)= \int_0^{+\infty}{G}_\text{b}\,(\epsilon,x,t)\,F\,(x+\mu)\,\text{d}x\,.
    \label{eq:partitionfunctionZb}
    \end{equation}

For any given initial nonequilibrium distribution $n_\mathrm{i}$, the NBDE including boundary conditions at the singularity can be solved through the nonlinear transformation Eq.\,(\ref{eq:Nformula}) and $Z\rightarrow  Z_\text{b}$. In the following section, we present the exact solution for $\theta$-function initial conditions that are appropriate for gluons in a relativistic heavy-ion collision through analytical calculations of the time-dependent partition function and its derivative.
\section{Time-dependent partition function for gluons}
We take the initial conditions for gluons in a relativistic heavy-ion collision as has been motivated e.g. in \cite{mue00} and used by \cite{jpb12},\cite{gw22} and others to be
\begin{equation}
	n_\mathrm{i}\,(\epsilon,t=0)=n_\text{i}^0\,\theta\,(1-\epsilon/Q_\mathrm{s})
	\label{inix}
\end{equation}
with an average initial occupation $n_\text{i}^0$ and a gluon saturation scale $Q_\text{s}$ that is of the order of 1 GeV. To obtain the generalized partition function, $F(x+\mu)$ must be evaluated from Eq.\,(\ref{fini}), resulting in
\begin{align}
    F(x+\mu)=\left\{\begin{array}{ll}
    \exp{\left(-\frac{n_\text{i}^1v}{2D}(x+\mu)\right)} &  x\leq Q_\text{s}-\mu \\
    \exp{\left(-\frac{v}{2D}(x+\mu+2n_\text{i}^0Q_\text{s})\right)} & x> Q_\text{s}-\mu\,,
    \end{array}\right.
\end{align}
where we have defined $n_\text{i}^1=2n_\text{i}^0+1$.
The generalized partition function and its derivative with respect to energy can now be calculated from Eq.\,(\ref{eq:partitionfunctionZb}). This was done in \cite{gw22} for inelastic gluon collisions with $\mu=0$ using the \texttt{NIntegrate} and \texttt{Derivative} routines of Mathematica. Here we aim to calculate the exact result fully analytically for arbitrary $\mu\le0$.

We start by separating the integral in the two domains on which $F$ is defined,
\begin{align}
\text{\small $Z(\epsilon,t)$}&\text{\small$=\int_0^{Q_\text{s}-\mu}\left[\exp{\left(-\frac{(\epsilon-\mu-x)^2}{4Dt}\right)}-\exp{\left(-\frac{(\epsilon-\mu+x)^2}{4Dt}\right)}\right]\exp{\left(-\frac{n_\text{i}^1v}{2D}(x+\mu)\right)}\,\text{d}x\nonumber$}\\
&\text{\small$+\int_{Q_\text{s}-\mu}^{\infty}\left[\exp{\left(-\frac{(\epsilon-\mu-x)^2}{4Dt}\right)}-\exp{\left(-\frac{(\epsilon-\mu+x)^2}{4Dt}\right)}\right]\exp{\left(-\frac{v}{2D}(x+\mu+2n_\text{i}^0Q_\text{s})\right)}\,\text{d}x$}\,.
\end{align}
A linear substitution $u=x+\mu$ yields
\begin{equation}
    Z(\epsilon,t)=\sum_{n=1}^4 I_n\,,
\label{eq2}    
\end{equation}
with the Integrals $I_n$ 
\begin{align}
    &I_1=\int_\mu^{Q_\text{s}}\exp{\left(-\frac{(\epsilon-u)^2}{4Dt}\right)} \exp{\left(-\frac{n_\text{i}^1v}{2D}u\right)}\,\text{d}u\,,\\
    &I_2=\int_\mu^{Q_\text{s}}-\exp{\left(-\frac{(\epsilon+u-2\mu)^2}{4Dt}\right)} \exp{\left(-\frac{n_\text{i}^1v}{2D}u\right)}\,\text{d}u\,,\\
    &I_3=\int_{Q_\text{s}}^{\infty}\exp{\left(-\frac{(\epsilon-u)^2}{4Dt}\right)} \exp{\left(-\frac{v}{2D}(u+2n_\text{i}^0Q_\text{s}\right)}\,\text{d}u\,,\\
    &I_4=\int_{Q_\text{s}}^{\infty}-\exp{\left(-\frac{(\epsilon+u-2\mu)^2}{4Dt}\right)} \exp{\left(-\frac{v}{2D}(u+2n_\text{i}^0Q_\text{s}\right)}\,\text{d}u\,.
\end{align}
These integrals are solved by rewriting the integrands as a Gaussians multiplied with factors that do not depend on the integration variable. The first integral $I_1$ becomes
\begin{align}
    I_1&=\int_{\mu}^{Q_\text{s}} \exp{\left(-\left(\frac{(\epsilon-u)}{\sqrt{4Dt}}-\frac{n_\text{i}^1v}{2}\sqrt{\frac{t}{D}}\right)^2\right)}\exp\left(-\frac{n_\text{i}^1v}{2D}\epsilon\right)\exp\left(\frac{(n_\text{i}^1)^2v^2}{4D}t\right)\,\text{d}u\,.
\end{align}
Substituting the term in the Gaussian
\begin{align}
    y=\frac{(\epsilon-u)}{\sqrt{4Dt}}-\frac{n_\text{i}^1v}{2}\sqrt{\frac{t}{D}}\hspace{2cm}\,\text{d}u=-\sqrt{4Dt}\,\text{d}y\,,
\end{align}
results in
\begin{align}
    I_1&=-\sqrt{4Dt}\exp\left(-\frac{n_\text{i}^1v}{2D}\epsilon\right)\exp\left(\frac{(n_\text{i}^1)^2v^2}{4D}t\right)\int_{x(u=\mu)}^{x(u=Q_\text{s})}\exp\left(-y^2\right)\text{d}y\nonumber\\
    &=-\sqrt{\pi Dt}\exp\left(-\frac{n_\text{i}^1v}{2D}\epsilon\right)\exp\left(\frac{(n_\text{i}^1)^2v^2}{4D}t\right)\nonumber\\
    &\times\left[\text{erf}\left(\frac{(\epsilon-Q_\text{s})}{\sqrt{4Dt}}-\frac{n_\text{i}^1v}{2}\sqrt{\frac{t}{D}}\right)-\text{erf}\left(\frac{(\epsilon-\mu)}{\sqrt{4Dt}}-\frac{n_\text{i}^1v}{2}\sqrt{\frac{t}{D}}\right)\right]\,.
    \end{align}
We proceed by solving the other integrals in a similar manner, yielding
\begin{align}
    I_2=&-\sqrt{\pi Dt}\exp\left(-\frac{n_\text{i}^1v}{2D}(\epsilon-2\mu)\right)\exp\left(\frac{(n_\text{i}^1)^2v^2}{4D}t\right)\nonumber\\
    &\times\left[\text{erf}\left(\frac{(\epsilon+Q_\text{s}-2\mu)}{\sqrt{4Dt}}+\frac{n_\text{i}^1v}{2}\sqrt{\frac{t}{D}}\right)
    -\text{erf}\left(\frac{(\epsilon-\mu)}{\sqrt{4Dt}}+\frac{n_\text{i}^1v}{2}\sqrt{\frac{t}{D}}\right)\right]\,,
\end{align}
\begin{align}
    I_3=\sqrt{\pi Dt}\exp\left(-\frac{v}{D}n_\text{i}^0Q_\text{s}\right)\exp\left(-\frac{v}{2D}\epsilon\right)\exp\left(\frac{v^2}{4D}t\right)\left[1+\text{erf}\left(\frac{(\epsilon-Q_\text{s})}{\sqrt{4Dt}}-\frac{v}{2}\sqrt{\frac{t}{D}}\right)\right]\,,
\end{align}
\begin{align}
    I_4=-\sqrt{\pi Dt}\exp\left(-\frac{v}{D}n_\text{i}^0Q_\text{s}\right)\exp\left(\frac{v}{2D}(\epsilon-2\mu)\right)\exp\left(\frac{v^2}{4D}t\right)\left[1-\text{erf}\left(\frac{(\epsilon+Q_\text{s}-2\mu)}{\sqrt{4Dt}}+\frac{v}{2}\sqrt{\frac{t}{D}}\right)\right].
\end{align}
We can now write $Z(\epsilon,t)$ as a sum of the integrals as in Eq.\,(\ref{eq2}). However, prefactors that do not depend on energy, namely $\sqrt{\pi Dt}$, are ignored because they vanish when calculating the occupation-number distribution through the logarithmic derivative of $Z(\epsilon,t)$. We thus obtain the generalized partition function for constant finite chemical potential $\mu\leq0$ as
\begin{align}
    Z(\epsilon,t)&=A(t)\Bigg\{\exp\left(-\frac{n_\text{i}^1v}{2D}\epsilon\right)\left[\text{erf}\left(\frac{\epsilon-\mu-n_\text{i}^1v\,t}{\sqrt{4Dt}}\right)-\text{erf}\left(\frac{\epsilon-Q_\text{s}-n_\text{i}^1v\,t}{\sqrt{4Dt}}\right)\right]\nonumber\\
    &\hspace{1.3cm}+\exp\left(\frac{n_\text{i}^1v}{2D}(\epsilon-2\mu)\right)\left[\text{erf}\left(\frac{\epsilon-\mu+n_\text{i}^1v\,t}{\sqrt{4Dt}}\right)-\text{erf}\left(\frac{\epsilon+Q_\text{s}-2\mu+n_\text{i}^1v\,t}{\sqrt{4Dt}}\right)\right]\Bigg\}\nonumber\\
    &+B(t)\Bigg\{\exp\left(-\frac{v}{2D}\epsilon\right)\left[1+\text{erf}\left(\frac{\epsilon-Q_\text{s}-v\,t}{\sqrt{4Dt}}\right)\right]\nonumber\\
    &\hspace{1.3cm}+\exp\left(\frac{v}{2D}(\epsilon-2\mu)\right)\left[-1+\text{erf}\left(\frac{\epsilon+Q_\text{s}-2\mu+v\,t}{\sqrt{4Dt}}\right)\right]\Bigg\}\,,
    \label{gpf}
\end{align}
with time-dependent auxiliary functions
\begin{align}
    &A(t)=\exp\left(\frac{(n_\text{i}^1)^2v^2}{4D}t\right)\\
    &B(t)=\exp\left(-\frac{v}{D}n_\text{i}^0Q_\text{s}\right)\exp\left(\frac{v^2}{4D}t\right)\,.
\label{eq_66}    
\end{align}
The derivative of the generalized partition function becomes
\begin{align}
   \frac{\partial Z(\epsilon,t)}{\partial\epsilon}= &\text{\small$ A(t)\Bigg\{-\frac{n_\text{i}^1v}{2D}\exp\left(-\frac{n_\text{i}^1v}{2D}\epsilon\right)\left[\text{erf}\left(\frac{\epsilon-\mu-n_\text{i}^1v\,t}{\sqrt{4Dt}}\right)-\text{erf}\left(\frac{\epsilon-Q_\text{s}-n_\text{i}^1v\,t}{\sqrt{4Dt}}\right)\right]\hspace{5cm}\nonumber$}\\
    &\text{\small$\hspace{0.9cm}+\frac{1}{\sqrt{\pi Dt}}\exp\left(-\frac{n_\text{i}^1v}{2D}\epsilon\right)\left[\exp\left(-\frac{\left(\epsilon-\mu-n_\text{i}^1v\,t\right)^2}{4Dt}\right)-\exp\left(-\frac{\left(\epsilon-Q_\text{s}-n_\text{i}^1v\,t\right)^2}{4Dt}\right)\right]\nonumber$}\\
    &\text{\small$\hspace{0.9cm}+\frac{n_\text{i}^1v}{2D}\exp\left(\frac{n_\text{i}^1v}{2D}(\epsilon-2\mu)\right)\left[\text{erf}\left(\frac{\epsilon-\mu+n_\text{i}^1v\,t}{\sqrt{4Dt}}\right)-
    \text{erf}\left(\frac{\epsilon+Q_\text{s}-2\mu+n_\text{i}^1v\,t}{\sqrt{4Dt}}\right)\right]\nonumber$}\\
    &\text{\small$\hspace{0.9cm}+\frac{1}{\sqrt{\pi Dt}}\exp\left(\frac{n_\text{i}^1v}{2D}(\epsilon-2\mu)\right)\left[\exp\left(-\frac{\left(\epsilon-\mu+n_\text{i}^1v\,t\right)^2}{4Dt}\right)-\exp\left(-\frac{\left(\epsilon+Q_\text{s}-2\mu+n_\text{i}^1v\,t\right)^2}{4Dt}\right)\right]\Bigg\}\nonumber$}\\
    +&\text{\small$B(t)\Bigg\{-\frac{v}{2D}\exp\left(-\frac{v}{2D}\epsilon\right)\left[1+\text{erf}\left(\frac{\epsilon-Q_\text{s}-v\,t}{\sqrt{4Dt}}\right)\right]\nonumber$}\\
    &\text{\small$\hspace{0.9cm}+\frac{1}{\sqrt{\pi Dt}}\exp\left(-\frac{v}{2D}\epsilon\right)\exp\left(-\frac{\left(\epsilon-Q_\text{s}-v\,t\right)^2}{4Dt}\right)\nonumber$}\\
    &\text{\small$\hspace{0.9cm}+\frac{v}{2D}\exp\left(\frac{v}{2D}(\epsilon-2\mu)\right)\left[-1+\text{erf}\left(\frac{\epsilon+Q_\text{s}-2\mu+v\,t}{\sqrt{4Dt}}\right)\right]\nonumber$}\\
    &\text{\small$\hspace{0.9cm}+\frac{1}{\sqrt{\pi Dt}}\exp\left(\frac{v}{2D}(\epsilon-2\mu)\right)\exp\left(-\frac{\left(\epsilon+Q_\text{s}-2\mu+v\,t\right)^2}{{4Dt}}\right)\Bigg\}$}\,.
    \label{derivative}
\end{align}
With Eqs.\,(\ref{gpf}),(\ref{derivative}), the -- within the model exact -- time-dependent gluon distribution function is now obtained from Eq.\,(\ref{eq:Nformula}).
\section{Limiting behaviour of the gluon distribution functions}
Before calculating time-dependent gluon distribution functions, we explicitly check the limiting behaviour of $n(\epsilon,t)$ for small and large times.
For $t\rightarrow 0$, the time-dependent distribution function must converge to the initial $\theta$-function, for $t\rightarrow \infty$ to the Bose--Einstein equlilibrium distribution. 
 With the conditions $Q_\text{s}>0$ and $\epsilon-\mu>0$, the time-dependent partition function becomes ($n_\text{i}^1=2n_\text{i}^0+1$ as defined above)
\begin{align}
    \lim_{t\to 0}Z(\epsilon,t)=\left\{\begin{array}{ll}
    2\exp{\left(-\frac{n_\text{i}^1v}{2D}\epsilon\right)} &  \epsilon< Q_\text{s} \\
    2\exp{\left(-\frac{v}{2D}\epsilon\right)}\exp{\left(-\frac{v}{D}n_\text{i}^0Q_\text{s}\right)} & \epsilon> Q_\text{s}\,.
    \end{array}\right.
\end{align}
Its derivative is given by
\begin{align}
    \lim_{t\to 0} \frac{\partial Z(\epsilon,t)}{\partial\epsilon}=\left\{\begin{array}{ll}
    -\frac{n_\text{i}^1v}{D}\exp{\left(-\frac{n_\text{i}^1v}{2D}\epsilon\right)} &  \epsilon< Q_\text{s} \\
    -\frac{v}{D}\exp{\left(-\frac{v}{2D}\epsilon\right)}\exp{\left(-\frac{v}{D}n_\text{i}^0Q_\text{s}\right)} & \epsilon> Q_\text{s}\,.
    \end{array}\right.
\end{align}
Hence, the occupation number in the limit $t\rightarrow 0$ is given by
\begin{align}
    \lim_{t\to 0}n(\epsilon,t)=-\frac{D}{v}\frac{\lim_{t\to 0} \frac{\partial Z(\epsilon,t)}{\partial\epsilon}}{\lim_{t\to 0}Z(\epsilon,t)}-\frac{1}{2}=\left\{\begin{array}{ll}
    n_\text{i}^0 &  \epsilon< Q_\text{s} \\
    0 & \epsilon> Q_\text{s}
    \end{array}\right.=n_\text{i}^0\Theta(Q_\text{s}-\epsilon)\,.
\end{align}
In the limit $t\rightarrow \infty$, we have $A(t)\rightarrow\infty$ in the time-dependent partition function, while the error functions cancel each other, resulting in $0\times\infty$. 
An analysis using l'Hôpital's rule shows, however, that
the terms with the prefactor $A(t)$ in the generalized partition function vanish in the limit $t\rightarrow\infty$, and we obtain
\begin{align}
    &\lim_{t\to \infty} Z(\epsilon,t)=\lim_{t\to \infty}2B(t)\left[\exp\left(-\frac{v}{2D}\epsilon\right)-\exp\left(\frac{v}{2D}(\epsilon-2\mu)\right)\right]\,,\\
    &\lim_{t\to \infty} \frac{\partial Z(\epsilon,t)}{\partial\epsilon}=\lim_{t\to \infty} -\frac{v}{D}B(t)\left[\exp\left(-\frac{v}{2D}\epsilon\right)+\exp\left(\frac{v}{2D}(\epsilon-2\mu)\right)\right]\,.
\end{align}
When calculating the occupation number, the time-dependent factor $B(t)$ cancels, resulting in
\begin{align}
    \lim_{t\to \infty}n(\epsilon,t)=-\frac{D}{v}\frac{\lim_{t\to 0} \frac{\partial Z(\epsilon,t)}{\partial\epsilon}}{\lim_{t\to 0}Z(\epsilon,t)}-\frac{1}{2}=\frac{1}{\exp\left(-\frac{v}{D}(\epsilon-\mu)\right)-1}\,.
\end{align}
Hence, the time-dependent occupation number converges to a Bose--Einstein distribution with temperature $T=-\frac{D}{v}$ and chemical potential $\mu$ as expected.
\section{Thermalization of massless gluons through inelastic scattering}

Inelastic gluon scatterings with $\mu=0$ mediate gluon splittings in the underoccupied case, and gluon fusion in overoccupied systems \cite{jpb12} where the initial gluon content is larger than that of the final equilibrium distribution, thus violating particle-number conservation.
Indeed, the NBDE solutions $n\,(\epsilon,t)$ for fixed chemical potential and density of single-particle states $g\,(\epsilon)=g_0\epsilon^2$ do not conserve the particle number \cite{gw22}
\begin{equation}
	N(t)=\int_0^\infty g\,(\epsilon)\,n\,(\epsilon,t)\,\text{d}\epsilon\ne \text{const}\,,
	\label{ntot}
\end{equation}
except for systems that have a critical initial occupation $n_\text{i}^0=0.154$. At this initial occupation, the particle-number density in equilibrium becomes equal to the one of a           $\theta$-function initial distribution, $N_\text{eq}^{\mu=0}/N=6\zeta(3)(15/4)^{3/4}/[(n_\text{i}^0)^{1/4}\pi^3]=1$  \cite{jpb12,blmt17}, thus defining the boundary from an under- to an overoccupied situation.
The solutions can therefore account for number-violating inelastic collisions with $\mu=0$ that lead to local statistical equilibrium in both, under- and overoccupied systems -- even though gluon splitting or fusion is not considered explicitly in the NBDE, but instead hidden in the transport coefficients.

To compute the time-dependent gluon distribution functions as given in Eq.\,(\ref{eq:Nformula}) using Eqs.\,(\ref{gpf}), (\ref{derivative}), the drift and diffusion coefficients for gluons enter as parameters since microscopic calculations are not available. Within the present model \cite{gw22}, their values are determined from the respective relations to the equilibrium temperature $T$, and the local bosonic equilibration time $\tau_\mathrm{eq}=4D/(9v^2)$ derived in Ref.\,\cite{gw18} at the gluon saturation scale $Q_\mathrm{s}$  as
\begin{equation}
 D=4\,T^2/(9\tau_\mathrm{eq})\,, \hspace{0.6cm} v=-4\,T/(9\tau_\mathrm{eq})\,.
 \label{Dv}
 \end{equation}
The relation between the transport coefficients and the local equilibration time arises from an asymptotic expansion of the error functions in the analytical solutions \cite{gw18} at the UV boundary $Q_\mathrm{s}$, whereas the fluctuation--dissipation relation $T=-D/v$ is a consequence of the equality of the stationary solution with the Bose--Einstein distribution as described before. 

For conserved energy density in the initial and the thermalized gluon distribution, the equilibrium temperature $T$ and the initial occupation were found to be related 
in Ref.\,\cite{jpb13} as $T=[15n_\text{i}^0/(4\pi^4)]^{1/4}Q_\mathrm{s}$, yielding $T\simeq443$\,MeV
for $Q_\mathrm{s}=1$\,GeV and $n_\text{i}^0= 1$, which we choose for convenience. This refers to an overoccupied system, as is typical for relativistic heavy-ion collisions at LHC energies. For example, the initial central temperature in $\sqrt{s_\text{NN}}=2.76$ TeV Pb-Pb collisions was determined to be $T_\text{ini}\simeq 480$ MeV \cite{hnw17} from bottomonium suppression in the quark-gluon plasma, corresponding to $n_\text{i}^0\simeq1.38$. 

Regarding the equilibration time for gluons, an experimental determination is not possible, one has to rely on the comparison between model calculations and data. From coupled 
kinetic equations \cite{fmr18} an upper limit of $\tau_\mathrm{eq}\leq 0.25$\,fm has been obtained, to be compared with an estimated freezeout time in central Pb-Pb collisions atLHC energies of $8-10$\,fm. However, these authors use a linear relaxation-time approximation in their numerical calculations of the collision term, such that a shorter value should be used in a nonlinear approach. Here we therefore take $\tau_\text{eq}=0.13$\,fm, and obtain the transport coefficients for an overoccupied system with $n_\text{i}^0=1$ from Eqs.\,({\ref{Dv}) as $D=0.67$\,GeV$^2$/fm and $v=-1.51$\,GeV/fm.

\begin{figure}
	\begin{center}
	\includegraphics[scale=0.3]{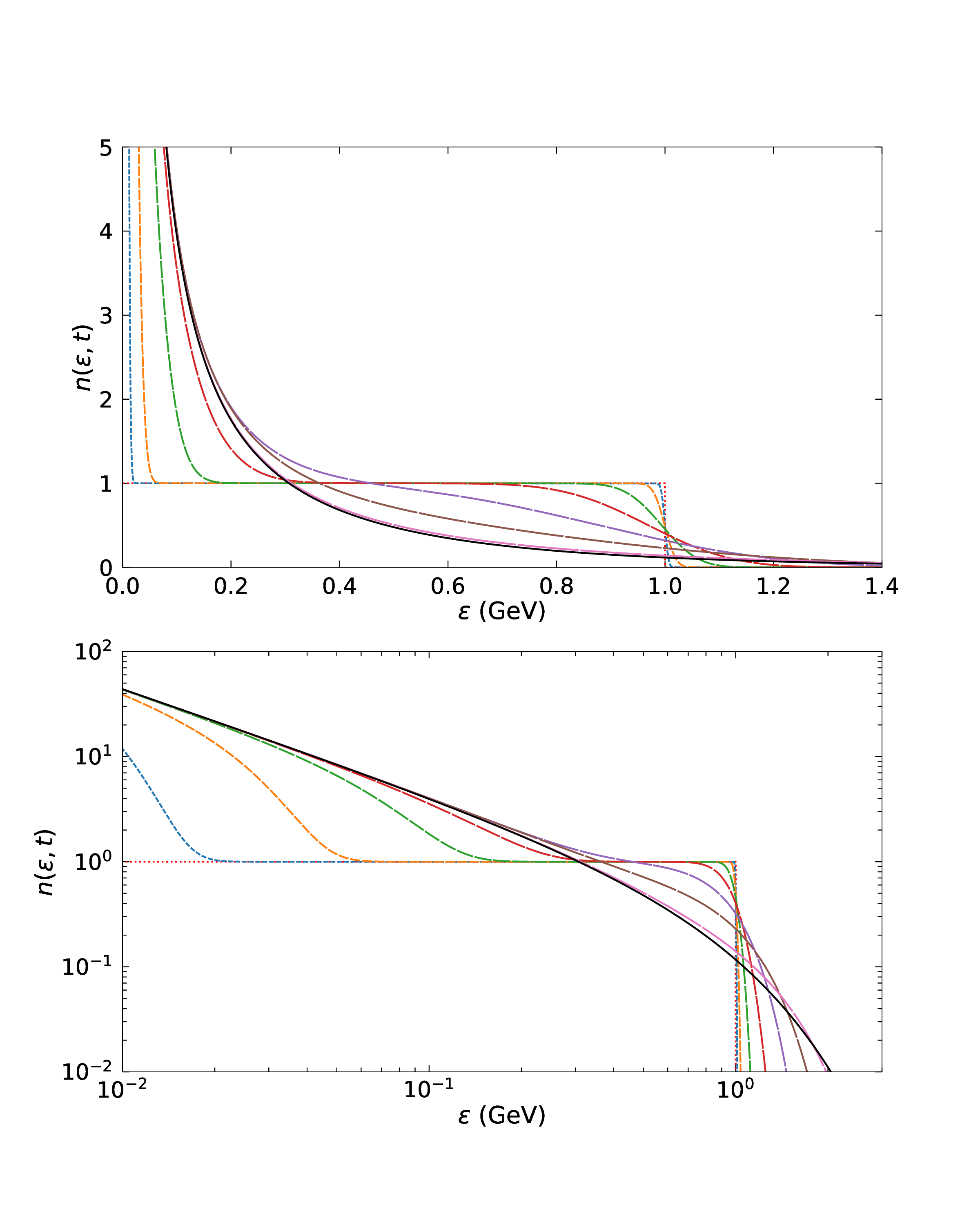}
	\caption{\label{fig1}
		Upper frame: Local thermalization of gluons in an overoccupied system with $n_\text{i}^0=1$ through inelastic collisions as represented by time-dependent exact solutions of the NBDE for constant transport coefficients and $\mu=0$. Starting from schematic colour-glass initial conditions 
		Eq.\,(\ref{inix}) in the cold system at $t=0$ (dotted line), a Bose--Einstein equilibrium distribution with temperature $T=443$\,MeV (solid curve) is approached. The transport coefficients are $D = 0.67$\,GeV$^2$/fm, 
$v=-1.51$\,GeV/fm. Time-dependent single-particle occupation-number distribution functions are shown at $t =2\times10^{-5}, 2\times10^{-4}, 2\times10^{-3}, 0.012, 0.04$ and $0.12$ and $0.4$\,fm (increasing dash lengths). 
Lower frame: Same results in a double-log plot.		
	}
	\end{center}
\end{figure}
\begin{figure}
	\centering
	\includegraphics[scale=0.3]{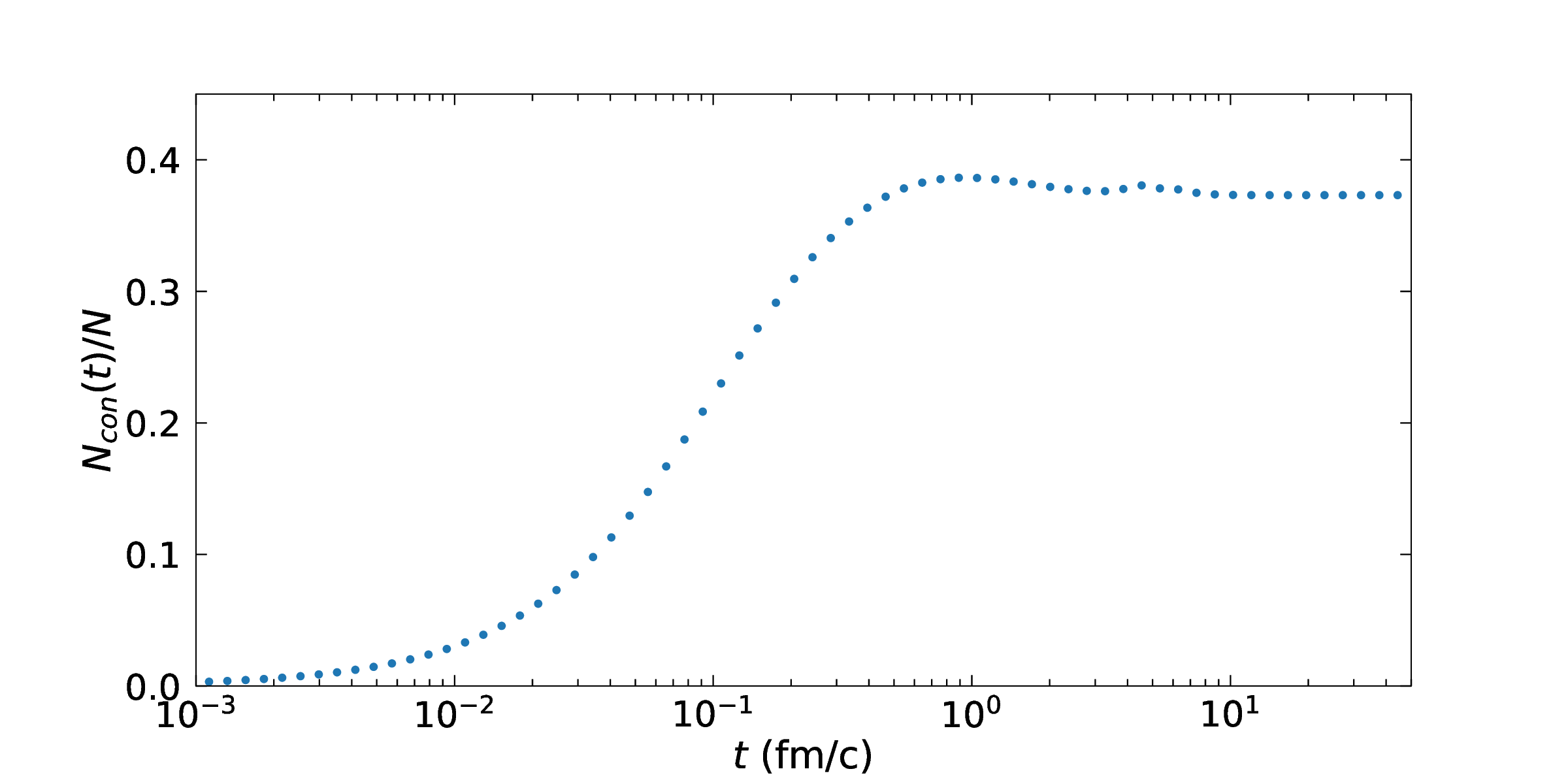}
	\caption{\label{fig2}
		Time-dependent number of condensed gluons $N_\text{con}(t)/N$ for an overoccupied system with  $n_\text{i}^0=1$. 
	}
\end{figure}

With these fixed average values for the drift and diffusion coefficients, the time evolution towards equilibrium via inelastic gluon collisions with $\mu=0$ can be calculated through the exact solution of the NBDE in the full momentum range. The results of the model calculation for gluon thermalization towards the Bose--Einstein equilibrium distribution (solid curve) are shown in Fig.\,\ref{fig1} for seven timesteps in a linear (top) and a double-log (bottom) scale. We have compared with our previous results for
inelastic gluon scattering, where the generalized partition function and its energy derivative had been obtained through numerical integration \cite{gw22}, albeit within the same nonlinear diffusion model. For any given timestep, these previous results are found to agree with our present exact analytical solution within at least eight decimal places. A discussion of inelastic collisions in under- and overoccupied systems has already been given in the previous work. The results were found to be consistent with QCD-based numerical calculations in the small-angle approximation that are, however, possible only in the infrared \cite{blmt17}.
\section{Gluon thermalization through elastic scattering}

 Based on QCD-inspired numerical solutions of Boltzmann-type transport equations for gluons, elastic collisions have been widely discussed in the literature \cite{jpb12,jpb13}, where the possibility of transient gluon-condensate formation in overoccupied systems was proposed. In the present context of a nonlinear diffusion equation, we can also consider elastic gluon collisions and time-dependent gluon condensate formation in overoccupied systems through the additional requirement of particle-number conservation. 
Technically, we supplement the explicit solution of the NBDE from section\,3 
 with a time-dependent boundary condition to account for particle-number conservation as function of time. For the generalized partition function, this implies $Z(\epsilon=\mu,t)=h(t)$ with an arbitrary function $h(t)\ge 0$, such that the transformation from the generalized partition function to the occupation-number distribution is well-defined. The detailed treatment is considered in the Appendix. The time-dependent boundary condition at $\epsilon=0$ is chosen such that the number of gluons
 in the thermal cloud (with $g_0=V/(2\pi^2)$ and a constant volume $V$ of the gluon distribution)
\begin{align}
    N_\text{th}=g_0\int_0^{\infty}\epsilon^2 n(\epsilon,t)\text{d}\epsilon
\label{numbercon}    
\end{align}
is conserved in time until there exists no solution to the NBDE. At this point, the gluons start to condense. The numerical calculation shows, however, that $N_\text{th}$ immediately decreases for homogeneous boundary conditions and hence, no solution exists such that $ N_\text{th}$ is conserved, and the time-dependent boundary condition corresponds to the homogeneous boundary condition. If considering a time-dependent parameter $\mu(t)$ \cite{kgw22} instead of a time-dependent boundary condition, the same result $\mu(t)=0$ for $t<0$ appears. Hence, the time-dependent distribution of the gluons consists of the analytical solution with vanishing chemical potential and homogeneous boundary conditions, and a discrete condensed state, which fulfils
\begin{align}
    N_\text{con}(t)=N-N_\text{th}(t)\,.
\end{align}
The time-dependent condensed particle numbers are shown in Fig.\,\ref{fig2} for $n_\text{i}^0=1$. We use the same transport coefficients $v, D$ as in the calculation for inelastic scattering because no detailed microscopic calculation is available and the results are meant to schematically illustrate the time-dependent physics of transient gluon condensation. Here, condensation starts immediately without a finite initiation time, which differs from the more detailed numerical model in \cite{blmt17} where the start of gluon condensation is delayed. In the time-dependent condensation of ultracold atoms, a finite initiation time has actually been found experimentally  \cite{kdg02}, but for gluons, a measurement is obviously impossible.

 At approximately $10^{-3}\,\si{fm}$, condensation in the NBDE approach with constant coefficients begins to rise significantly until approximately $1\,\si{fm}$. For times larger than $1\,\si{fm}$, the number of condensed gluons drops slightly until it reaches the equilibrium distribution, differing from the physical expectation of a monotonically increasing function converging to the equilibrium value as an upper bound. This upper bound is given by the initial particle number minus the particle number in equilibrium as \cite{blmt17}
\begin{align}
    N^{\text{eq}}_{\text{con}}=N-N^{\text{eq}}_{\text{th}}=g_0 Q_\text{s}^3\left[\frac{1}{3}n_\text{i}^0-2\zeta(3)\left(\frac{15}{4\pi^4}n_\text{i}^0\right)^{3/4}\right]\,.
\label{eq_condensed}    
\end{align} 
 The total gluon number from the initial $\theta$-function distribution is $N=g_0n_\text{i}^0 Q_\text{s}^3/3$, such that $ N^{\text{eq}}_{\text{con}}/N=0.373$ for $n_\text{i}^0=1$. 
This agrees with the stationary NBDE  limit shown in  Fig.\,\ref{fig2}. The slight overshoot of the equilibrium value is a consequence of the constancy of the transport coefficients $v, D$ in the analytical model: The large constant drift towards the IR causes thermalization in the UV region to be too slow, such that at large times $t\simeq 1$\,fm the Bose--Einstein limit is reached in the IR, but not yet in the UV, resulting in an unbalanced energy-dependent thermalization. As a consequence, the number of gluons in the thermal cloud $N_\text{th}(t)$ is too small near the saturation region, and $N_\text{con}(t)/N$ overshoots the equilibrium limit. This deficiency could be remedied by introducing energy-dependent transport coefficients, but then, the NBDE cannot be solved analytically anymore.

To which extent the approximation of constant coefficients is indeed satisfied for gluons of Quantum Chromodynamics is difficult to determine. At present, a justification can be drawn from the fact that the constant-coefficient model reproduces qualitative -- and to some extent, quantitative -- features of the behaviour of the gluon distributions that are obtained in numerical solutions of more sophisticated nonlinear gluon transport equations such as Ref.\,\cite{blmt17}. These calculations are, however, only valid in the infrared region, and make use of the small-angle approximation of gluon scattering. A detailed comparison of their numerical results with the same initial $\theta$-function gluon distribution and $n_\text{i}^0=1$ is in remarkably good agreement with our schematic constant-coefficient NBDE approach in the IR once a realistic physical timescale is introduced into their model. This holds up to about $\epsilon=300$ MeV, when their numerical approach ceases to work.

It will be interesting to compare the present constant-coefficient approach with the numerical solution of the full NBDE (\ref{nbde}) for physically reasonable energy dependencies of both, drift- and diffusion coefficients, keeping $\lim_{t\rightarrow \infty}[-v\,(\epsilon,t)/D\,(\epsilon,t)] \equiv 1/T$ as before. When using the NBDE to account for the thermalization of ultracold atoms and time-dependent Bose--Einstein condensate formation, we have already obtained numerical solutions for energy-dependent transport coefficients \cite{lgw24}: With $D(\epsilon)=\alpha\epsilon\exp(-\beta\epsilon)$ and a mean value $\langle D(\epsilon)\rangle$ matching the constant coefficient for suitably chosen $\alpha,\beta$, 
the time evolution of the single-particle occupation-number distribution turns out to be rather similar to the constant-coefficient case. When calculating time-dependent condensate formation, both cases are in reasonable agreement with recent time-dependent condensate formation data for $^{39}$K at various scattering lengths. However, thermalization in the UV is reached faster for energy-dependent coefficients -- which actually prevents the slight overshoot of the time-dependent condensate fraction above the thermal limit that was observed for gluons in Fig.\,\ref{fig2}, and is also present -- but less pronounced -- in cold atoms for constant coefficients. We are preparing a similar calculation with energy-dependent coefficients for gluons, where the initial condition is significantly different from the cold-atom case.

\section{Conclusions}

We have developed an exact analytical solution of the nonlinear boson diffusion equation for inelastic and elastic gluon scattering in relativistic heavy-ion collisions at LHC energies.
With $\theta$-function initial conditions, gluons up to the saturation momentum $Q_\text{s}\simeq1$\,GeV are quickly thermalized in the initial stages of the collision that last less than 1\,fm, whereas the typical interaction (freezeout) time in a central Pb-Pb collision at LHC energies is $8-10$\,fm. Using constant drift and diffusion coefficients and a nonlinear transformation that has been developed earlier, the NBDE has been solved with the full consideration of boundary conditions at the singularity occuring at the value of the chemical potential. The exact solution agrees to at least eight decimal places with an earlier calculation for inelastic scattering where the time-dependent partition function and its derivative
that enter the nonlinear transformation had been computed numerically. The analytic calculation of the time-dependent partition function and its derivative -- and thus, the solution of the NBDE through the nonlinear transformation -- is the main advance of this work.

We have also calculated transient gluon-condensate formation in elastic collision using the additional condition of particle-number conservation with a corresponding time-dependent
boundary condition. In the simplified model with constant transport coefficients, the condensate initiation time vanishes. For reasonable values of drift and diffusion, condensate formation starts significantly at $10^{-2}$\,fm and rises towards the statistical equilibrium limit in a typical overoccupied system of gluons until about $1$\,fm. As a consequence of the constant coefficients, a slight overshoot of the equilibrium value occurs. It could be remedied using energy-dependent coefficients, but then the
analytical treatment of the nonlinear problem would become impossible.
In such a more detailed model for nonlinear diffusion of gluons, one may also expect a finite value for the condensate initiation time, as has been found in 
the numerical model of Blaizot et al., yielding the conclusion that transient gluon condensation is unlikely to occur because the timescale for inelastic collisions is shorter.

\bigbreak
\noindent
\noindent
  \renewcommand{\theequation}{A-\arabic{equation}}
  \setcounter{equation}{0}  
  \section*{Appendix: Time-dependent boundary conditions}
Time-dependent boundary conditions have been considered in the literature, see e.g. \cite{er18}. Here, we specifically
consider the boundary-value problem on $\epsilon \in[0,\infty)$ with a homogeneous initial condition $Z(\epsilon,t=0)=0$ and an inhomogeneous boundary condition $Z(\epsilon=0,t)=h(t)$. The solution $u$ is the convolution of $-2D\partial_xG(x,t)$ and the boundary condition $h(t)$ with respect to $t$, where $G(\epsilon,t)$ is the free Green's function introduced in Section\,3 as $G_\text{free}(\epsilon,x,t)$ for $x=0$. This solution reads
\begin{align}
u(\epsilon,t)=\int_0^{t}\underbrace{\frac{\epsilon}{\sqrt{4\pi D(t-s)^3}}\exp\left(-\frac{\epsilon^2}{4D(t-s)}\right)}_{=g(s)}h(s)\text{d}s\,,
\label{eq_2000}    
\end{align}
which is valid $\forall \epsilon>0$. One can check that this solution fulfils the linear diffusion (heat) equation and converges to the boundary condition $h(t)$ for $\epsilon\rightarrow 0$. 

Given an initial condition and a time-dependent boundary condition, the general solution of the generalized partition function and hence the occupation-number distribution on           $\epsilon\in[0,\infty)$ can be calculated. The time-dependent boundary condition is not explicitly known when considering elastic gluon scattering, but the number of gluons is known to be conserved. This corresponds to an implicit condition for the boundary condition, which has the form
\begin{align}
    S[n(\epsilon,t)]:=g_0\int_0^{\infty}\epsilon^2n(\epsilon,t)d\epsilon-N=0\,,
\end{align}
where $N$ is the total particle number, which is a constant positive real number.
The problem can be solved numerically for discrete times with the following technique. Approximating $h(t)$ as a step function
\begin{align}
    h(t)=\left\{\begin{array}{llll}
    h(\Delta t) &  t\leq  \Delta t \\
    h(2\Delta t) & \Delta t<t\leq 2\Delta t\\
    ... & ...\\
    h(n\Delta t) & (n-1)\Delta<t\leq n\Delta t
    \end{array}\right.
\end{align}
with $t=n\Delta t$, and rewriting the integral in Eq.\,(\ref{eq_2000}) yields
\begin{align}
    u(\epsilon,t)=\int_0^tg(s)h(s)\text{d}s=\lim_{n \to \infty}\sum_{k=1}^{n}h(k\Delta t)\int_{(k-1)\Delta t}^{k\Delta t}g(s)\text{d}s\,.
\label{eq_3004}    
\end{align}
The remaining integral in Eq.\,(\ref{eq_3004}) can be solved analytically by substituting 
\begin{align}
    y=\frac{4D}{\epsilon^2}(t-s)\hspace{2cm}\text{d}y=-\frac{4D}{\epsilon^2}\text{d}s\,,
\end{align}
resulting in
\begin{align}
    \int_{(k-1)\Delta t}^{k\Delta t}\frac{\epsilon}{\sqrt{4 D(t-s)^3}}\exp\left(-\frac{\epsilon^2}{4D(t-s)}\right)\text{d}s=-\int_{y((k-1)\Delta t)}^{y(k\Delta t)}\frac{1}{y^{3/2}}\exp\left(-\frac{1}{y}\right)\text{d}y\,.
\end{align}
Substituting again 
\begin{align}
    \xi=\frac{1}{\sqrt{y}}\hspace{2cm}\text{d}\xi=-\frac{1}{2}\frac{1}{y^{3/2}}\text{d}y\,,
\end{align}
yields
\begin{align}
    &\int_{(k-1)\Delta t}^{k\Delta t}g(s)\text{d}s=2\int_{\xi(y((k-1)
    \Delta t))}^{\xi(y(k\Delta t))} \exp(-\xi^2)\text{d}\xi\nonumber\\
    &=\sqrt{\pi}\left[\text{erf}\left(\frac{\epsilon}{\sqrt{4D(n-k)\Delta t}}\right)-\text{erf}\left(\frac{\epsilon}{\sqrt{4D(n-k+1)\Delta t}}\right)\right]\,.
\end{align}
Therefore, $u(\epsilon,t)$ can be written as
\begin{align}
    u(\epsilon,t)=\sqrt{\pi}\lim_{n \to \infty}\sum_{k=1}^{n}h(k\Delta t)\times\left\{\begin{array}{ll}
    \text{erf}\left(\frac{\epsilon}{\sqrt{4D(n-k)\Delta t}}\right)-\text{erf}\left(\frac{\epsilon}{\sqrt{4D(n-k+1)\Delta t}}\right)  & k<n \\
    1-\text{erf}\left(\frac{\epsilon}{\sqrt{4D(t-k+1)\Delta t}}\right) & k=n\,,
    \end{array}\right.
\end{align}
where one has to distinguish between $k<n$ and $k=n$, in order that the expression is well defined. Its derivative with respect to $\epsilon$ becomes
\begin{align}
    \sqrt{\frac{\pi}{4Dt}}\lim_{n \to \infty}\sum_{k=1}^{n}h(k\Delta t)\times\left\{\begin{array}{ll}
       \frac{1}{\sqrt{n-k}}\exp\left(-\frac{\epsilon^2}{4D(n-k)\Delta t}\right)-\frac{1}{\sqrt{n-k+1}}\exp\left(-\frac{\epsilon^2}{4D(n-k+1)\Delta t}\right)  & k<n \\
        -\frac{1}{\sqrt{n-k+1}}\exp\left(-\frac{\epsilon^2}{4D(n-k+1)\Delta t}\right) & k=n\,.
    \end{array}\right.
\end{align}
The time-dependent occupation number is then given by
\begin{align}
    n(\epsilon,t)=-\frac{D}{v}\frac{\partial}{\partial \epsilon}\ln(w(\epsilon,t)+u(\epsilon,t))-\frac{1}{2}\,.
\end{align}
Under a given implicit condition like particle number conservation, $h(t)$ can be calculated numerically for discrete time steps $\Delta t$. The first time step $t=\Delta t$ fixes $h(t<\Delta t)$ through imposing $S[n(\epsilon,\Delta t)]=0$, which is then used to calculate $h$ in the next time step $t=2\Delta t$. This step fixes $h(t<2\Delta t)$. This algorithm is repeated until reaching $t=n\Delta t$ or until there exists no solution such that $S[n(\epsilon,t)]=0$, which is the case for an overoccupied system.  In the limit of infinite time steps this technique is exact. In reality, the number of time steps is limited by the performance of the numerical treatment.\\


\section*{References}
\bibliographystyle{iopart-num}
\bibliography{gw_24}
\end{document}

%% file: macros.tex
\NewDocumentCommand{\tup}{smo}{\IfNoValueTF{#3}{\bm{#2}}{\IfBooleanTF{#1}{{#2}_{#3}}{{#2}^{#3}}}}
\DeclarePairedDelimiterX{\tupdelims}[1]{\lparen}{\rparen}{\cramped{#1}}


\DeclarePairedDelimiterX{\setdelims}[1]{\lbrace}{\rbrace}{\cramped{#1}}

\DeclarePairedDelimiterXPP{\func}[2]{#1}{\lparen}{\rparen}{}{\cramped{#2}}

\NewDocumentCommand{\drift}{om}{\tup{\mu}[#1]_{#2}}
\NewDocumentCommand{\diffusion}{om}{\tup{\sigma}[#1]_{#2}}
\NewDocumentCommand{\diffusivity}{om}{\tup{D}[#1]_{#2}}

\NewDocumentCommand{\rap}{o}{\tup{\xi}[#1]}
\NewDocumentCommand{\Rap}{o}{\tup{\Xi}[#1]}
